# Single-beam noise characteristics of quantum correlated twin beams

**Yun Zhang\*, Katsuyuki Kasai, and Kazuhiro Hayasaka**

Kansai Advanced Research Center, Communications Research Laboratory

588-2 Iwaoka, Nishi-ku, Kobe, 651-2492 Japan

**Abstract**. We investigated the intensity noise spectra of the single beam of a pump-enhanced continuous-wave (cw) optical parametric oscillator (OPO), which was used to generate quantum correlated twin beams, as a function of the pump power. With a triply (pump-, signal-, and idler-) resonant cavity, the oscillation threshold of our OPO was about 8.5±1.3 mW and the measured slope conversion efficiency was 0.72±0.02. A twin beams with power of 240 mW were generated at pump power of 350 mW. The relaxation oscillation frequencies, which depend on the pump power, were observed when the pump power of OPO was from 12.5 mW to 28 mW. The experimental results confirm the predicted increase in OPO relaxation frequency with pump power. Squeezing of the single beam intensity was for the first time inferred experimentally by exploiting nature of quantum noise dependent on loss.

# 1 Introduction

Continuous-wave (CW) Optical Parametric Oscillators (OPOs) are efficient and widely tunable sources of coherent light [1] and nonclassical states [2-11]. Degenerate cw OPOs have been used to produce single- mode quadrature squeezed light of a single mode [2]. The quantum correlated twin beams with a squeezed intensity difference fluctuation and two-mode quadrature squeezed vacuum state of light have been generated from nondegenerate OPOs respectively operating below and above the oscillation threshold [3,4]. The entangled state, with quantum entanglement between the quadrature phase amplitudes of spatially separated signal and idler beams, has been experimentally generated from a nondegenerate OPO operating below the oscillation threshold [5]. Twin beams generated by OPOs have been observed in a number of experiments on type II phase matching devices since 1987 [3, 4, 6-11]. Since then twin beams has been applied to high-sensitivity and quantum nondemolition measurements [12], as well as to high-sensitivity spectroscopy [13,14].

Pump-enhanced OPOs, also referred to as triply resonant OPOs, are significant twin beams sources. Compared with single- and double-resonant OPOs, they have very low thresholds and provide stable single-frequency output. Low-threshold OPOs are of interest for investigating the quantum properties of beams generated in the high pump power regime (i.e. for pump powers of at least four times the threshold), especially the single-beam noise characteristics of twin beams. Up to now, most of the reported measurements of cw twin beams correlation have been carried out with low pump power (for pump powers low than four times the threshold), and most have focused only on the correlation between the signal and idler beams. Furthermore, the output power of the twin beams of these devices is low, about several dozen milli-watts. All of these were limited the applications of the laser-like twin beams. The high pump regime [11] is of critical interest for using OPOs as sources of cw tunable laser-like radiation for making high-sensitivity spectroscopic measurements. In addition to tunable character, these devices offer the unique potential of providing high-power twin beams and of investigating the single-beam noise characteristics of twin beams.

Pump-enhanced OPOs exhibit several unique properties. The intensity noise spectra of the output single beam of a cw triply resonant OPO is similar to the noise spectra of a lser oscillator, which exhibits relaxation oscillation. Recently, Lee et al. [15] observed relaxation oscillations in the intensity noise spectrum of the single beam emitted by a non-degenerate OPO pumped by a diode laser, and Porzio et al. [11] observed relaxation oscillations in the single beam spectra of twin beams output from triply resonant OPO for pump powers up to about fifteen times the threshold value. The relaxation oscillation frequency shifts to higher values as the pump power is increased. Working with our OPO, we have observed a similar feature in the single-beam spectra. Another important property of triply resonant OPOs is that the intensity noise in a single beam of twin beams can be reduced below the shot noise limit. In theory, it was predicted that this intensity noise falls below the shot noise limit when the pump power is above four times the oscillation threshold in a bandwidth of the order of the cavity bandwidth [16]. In the ideal case, the intensity noise in the single beam of twin beams is reduced to half of the shot noise limit at very high pump powers. Recently, Porzio et al. have developed a complete model for explaining the noise source that influence an OPO [7]. In this paper we will study the intensity difference noise between the signal and idler beams and intensity noise of single beam at high pump range in detail.

The purpose of the present paper is to present the single-beam noise characteristics of twin beams generated by a triply resonant OPO. The relaxation oscillation frequencies, which depend on the pump power, were observed when the pump power of the OPO was from 12.5 mW to 28 mW. The experimental results confirm the predicted increase in OPO relaxation frequency with pump power. Squeezing of the single beam intensity was also experimentally studied for the pump powers of more than 60 mW when the extra noise was canceled. The paper is organized as follows. Section 2 describes the experimental setup. Section 3 describes and discusses the experimental results. The relaxation oscillation of our OPO and the intensity noise on the single beam of twin beams are discussed in detail. Section 4 presents our conclusions.

## 2 Experimental setup

The experimental setup is shown schematically in Fig. 1. The OPO was the same pump-enhanced signal and idler-resonant (triply resonant) OPO as was used in [8, 9] to generate the quantum correlated twin beams. A diode-pumped Nd:YAG (Lightwave Electronics, Model 142) serves as the pump source of the system. The laser delivered a maximum output power of 400 mW at a fixed wavelength of 532 nm, of which 350 mW was available in front of the OPO. A half-wave plate (P1) and a polarizing beam splitter (PBS0) were used to adjust the power in front of the OPO. The twin beams were produced when the OPO was operating above the threshold. The signal and idler beams of the twin beams were cross-polarized. To measure the single beam's noise characteristic, the twin beams were separated into the signal and idler beams by a polarizing beam splitter (PBS1) according to their polarization. The intensity noise of the signal beam was detected in a balanced homodyne detector [17]. The detection apparatus was composed of a half-wave plate (P2), polarizing beam splitter (PBS2) and two identical high quantum efficiency photodiodes (D1 and D2 InGaAs, Epitaxx ETX500) matched to equal transimpedence low noise amplifiers. The half-wave plate and polarizing beam splitter acted to split the beam into two beams with equal intensity. The ac photocurrents of the balanced detector were combined in a hybrid junction to generate the sum and difference currents $i_1$ and $i_2$. These currents were input to spectrum analyzers that recorded the noise powers. The sum signal was a measure of the intensity noise of the beam, while the difference signal gave the shot noise level [17].

The half-wave plate (P3, enclosed by the dashed line in Fig. 1) and PBS1 were removed to measure the correlation between the signal and idler beams. As shown by Heidmann [3], the half-wave plate (P2), PBS2, and two detectors composed the detection system for the twin beams. To measure the noise of single beam, P3 and PBS1 were inserted. In this case, the angle between the axes of the plate (P2) and of the polarizer (PBS2) was fixed at 22.5° and the system of P2 and PBS2 acts like a usual 50% beam splitter. When the angle between the axes of the plate (P3) and of the polarizer (PBS1) is 0°, the transmission light of PBS1 is the signal beam; and when the angle between the axes of the plate (P3) and of the polarizer (PBS1) is 22.5°, transmission light of PBS1 is the mixed light of the signal and idler beam. In this way, the intensity noises of signal beams and intensity noises of mixed beam between signal and idler beams can be recorded, respectively. The process of detection will be described in section 3.4.

## 3 Experimental results

### 3.1 Steady operation

A calibrated thermal power meter was employed for both the pump power ($P_p$) and output power ($P_{out}$) measurements. Figure 2 shows the results of measurements made over a range of pump powers, from 18 to 350 mW. The input power was varied by rotating the half-wave plate (P1) of the power adjustment system. In particular, at the pump power of 350 mW, an output power of 240 mW was obtained. From these measured powers, we computed the conversion efficiency $\eta = P_{out}/P_p$, obtaining $0.68 \pm 0.04$ for the frequency down conversion from 532 nm to 1064 nm. The linear fitting is also shown in Fig. 2 (dot line), which yields an OPO threshold of $P_{th} = 9.0 \pm 1.8$ mW and a slope conversion efficiency of $0.72 \pm 0.02$.

Solving the equations of evolution in the cavity and using the boundary condition at the output coupler, the output power can be expressed as the square-root of the pump power or as a function of the threshold factor [16],

$$P_{out} = 2\varepsilon(\sqrt{P_{th}P_p} - P_{th}) \text{ or } p_{out} = 2\varepsilon(\sqrt{s} - 1) \qquad (1)$$

where $p_{out} \equiv P_{out}/P_{th}$ and $s \equiv P_p/P_{th}$ are the normalized output power and normalized pump power, respectively. The constant parameter, $\varepsilon$, represents the power slope efficiency of the output beam at the threshold. Equation (1) was used to fit the experiment data, to determine threshold $P_{th}$, and thus to express the measured pump powers with the threshold factors. We used a two-parameter least squares fitting of Eq. (1) to the experimental data in Fig. 2, where $\varepsilon$ and $P_{th}$ were used as the fitting parameters (solid line). From the good agreement (inset in Fig. 2) in the low pump power range (pump powers less than 13 times the pump threshold), the fit yielded a pump power at a threshold of $P_{th} = 8.5 \pm 1.3$ mW, which is near the threshold value of the linear fit of all experimental data, and s power slope efficiency of $\varepsilon = 1.2 \pm 0.1$. Unfortunately, the output power fell short of the theoretically expected value in the high pump range. In Fig. 2, the output power is higher than the expected value. Possible reasons for this discrepancy are: increased subharmonic loss and decreased pump loss because of the up-conversion, absorption of the OPO fields in the crystal, or thermally induced changes in the crystal's refractive index, etc. All these reasons will induce changes in the pump threshold and power slope efficiency, and in turn, in the output power of the OPO. These issues will have to be carefully addressed in future work.

Note that during each increment of the pump power, we observed several axial mode hops of the subharmonic wave, which yielded different output powers. Such mode hops are probably induced by absorption of the OPO fields in the crystal. This leads to thermally induced changes in the crystal's refractive index and spectral clustering [1, 18]. The difference in output power can be interpreted as a varying threshold or threshold factor. Reliable operation of the OPO at a chosen threshold factor thus required a careful adjustment of the cavity length. For a reproducible measurement of output power vs. pump power, we fine-tuned the cavity length during cw-OPO resonant operation. This was done by applying a manually adjustable offset voltage to the PZT. For each pump power setting, the voltage was adjusted to produce the maximum output power.

## *3.2 Intensity correlation measurements*

The measured intensity difference spectrum of the twin beams, which is generated by a non-degenerate OPO, is expressed as

$$S(\Omega) = 1 - \{\eta_e \eta_d /[1+(\Omega/\Gamma)^2]\}, \qquad (2)$$

where $\Omega$ is the analysis frequency, $\Gamma$ is the cavity linewidth, and $\eta_e = T_2/(T_2 + T_1 + \alpha)$ is the cavity escape efficiency, with $T_2 = 5\%$ being the transmission of the output mirror, $T_1 < 0.1\%$ the transmission of the high reflectivity facet of the crystal, and $\alpha < 0.3\%$ the extra losses. The total detection efficiency, $\eta_d$, is about 90%. The detection efficiencies in our experiment ware measured with ~92% photodetector quantum efficiency and ~98% propagation efficiency.

Figure 3 shows the best intensity difference noise reduction of the twin beams generated by our OPO. The frequency range in the figure is from 1 to 50 MHz. The maximum noise reduction was $-7.2 \pm 0.3$ dB (80%) below the shot-noise limit, at an analysis frequency of 3 MHz. The total output power of 26 mW (13 mW per beam) was obtained when the pump power was about 40 mW (about four times the threshold). The values predicted by Eq. (2) (solid curve) with the experimentally measured parameters ($\eta_e$, $\eta_d$ and $\Gamma$) are shown for comparison. The experiment results agree with the theory very well. The output power of the twin beams increased with pump power. Because of the detection saturation of the detector, we did not detect the best correlation between the signal and idler mode. However, part of the twin beams was measured with our detection system; the correlation between signal and idler mode was always retained (the correlation between signal and idler decreased due to the low detection efficiency), even with the output power of 240 mW. The noise reduction of an attenuated twin beams with power of 100mW is shown in inset of Fig. 3. The noise reduction was $-1.2 \pm 0.3$ dB below the shot-noise limit, because the attenuation was induced. This means that the twin beams with power of 240 mW were generated with our OPO.

## 3.3 Relaxation oscillation

The intensity noise in a single beam of the OPO output was measured by using balanced detection to detect the single beam of twin beams. Examples of the recorded normalized single-beam intensity noise as a function of noise frequency are shown in Fig. 4 for the OPO operating at the pump powers of 12.5, 18, 23, and 28 mW. At high pump powers, the noise spectra were different and clearly showed a broad noise peak with a non-zero center frequency, which is called the relaxation oscillation frequency. This frequency shifts to higher values as the pump power is increased. However, an unpredicted experimental result was observed when the pump threshold factor increased further. Figure 5 shows the normalized intensity noise in a single beam versus frequency for pump powers from 28 to 55 mW. The intensity noise of signal beams decreases with increasing pump power. This is different from the previous case [11, 15]. The intensity noise decreased with increasing pump power, indicating the inherent noise character of the single beam generated by an OPO. This behavior is like that predicted by Fabre [16] and was recently studied theoretically by Porzio in detail [19]. The reduction in the single beam intensity noise indicates a possibility of using our triply resonant OPO to generate sub-Poissonian light at high pump powers.

*3.4 Single beam noise spectra*

Except for the relaxation oscillation, it is more attractive to generate the sub-Poissonian light from an OPO operating above the threshold. In the absence of pump excess, squeezing should appear in the single-beam spectrum of a perfectly triply resonant OPO for pump powers well above four times the threshold value [16, 19]. To reach such a regime it is necessary to lower the threshold of the OPO and shift the relaxation oscillation frequency to a high value as quickly as possible while increasing the pump power. In our OPO, because of the large pump excess noise and tail of relaxation oscillation, the squeezed single beam of the twin beams was not obtained. However, we did indicate for the first time the squeezing of the single beam by exploiting nature of quantum noise dependent on loss when the pump power was more than four times the pump threshold. In order to indicate the squeezing, we separated the detected intensity noise into classical excess noise and quantum noise in the single beam as follows:

$$V_d = V_{ex} + V_q, \qquad (3)$$

where $V_d$ is the detected noise, $V_{ex}$ is the classical excess noise, which is above the shot noise limit, and $V_q$ is quantum noise, which is either below the shot noise limit or not. If the noise is below the shot noise limit, it will be very sensitive to loss and any loss will induce the extra noise (shot noise) into the system, i.e., $V_{out} = \mu V_{in} + (1-\mu)V_{SNL}$ where $V_{SNL}$ is the shot noise and $\mu$ is the loss efficient. However if the noise is classical noise, there will be no extra noise (shot noise) added to system due to the loss, i.e., $V_{out} = \mu V_{in}$. Thus it is possible to distinguish the quantum noise and classical noise by adding the loss to the system. In experiment, the measurement process consisted of two steps. In the first step, the intensity noise of the single beam without loss (the intensity noise of signal beam) was measured using a homodyne detector. This means the sum of excess noise and quantum noise was recorded. In the second step, the intensity noise of mixed light (it was realized by rotating the half wave plate P3), which was combined with half of the signal and half of the idler, was measured with the homodyne detector. In this case, the sum of excess noise and quantum noise after loss (loss efficiency is 1/2) was recorded. Comparing with these two noise powers, the quantum noise of a single beam can be found. Supposing the quantum noise $V_q = S_s V_{SNL}$, where $S_s$ is squeezing, the sum of excess noise and quantum noise after loss can be written as $V_{d(\mu)} = \mu V_{ex} + \mu[\mu S_s + (1-\mu)]V_{SNL}$. Thus the detected noises in our first and second step measurement are $V_{d(I)} = V_{ex} + S_s V_{SNL}$ and $V_{d(II)} = V_{d(0.5)} + V_{d(0.5)} = V_{ex} + \frac{1}{2} S_s V_{SNL} + \frac{1}{2} V_{SNL}$, respectively, when the system is balanced just in case of our OPO. The squeezing can be written

$$S_s = 1 - 2(\frac{V_{d(II)}}{V_{SNL}} - \frac{V_{d(I)}}{V_{SNL}}). \tag{4}$$

Thus, the difference between two measurements indicates squeezing of the quantum noise directly. Examples of intensity noise as a function of analysis frequency are shown in Fig. 6 (a)-(d) for the OPO operating at different pump powers. When the pump power was below four times the oscillation threshold, there was no difference in the intensity noise between with and without loss. This indicates that the quantum noise of the single beam is above or equals the shot noise limit. At high pump powers (more than four times of the threshold), (b), (c) and (d), the intensity noise powers with and without loss were different and clearly show squeezing of the single beam. To our knowledge, the squeezing on single beam output from OPO has never been observed and indicated in experiment before. To confirm that this is the case, we compared our experimental results with the theoretical prediction. In the absence of pump excess, the spectrum of the single beam can be written as [16]

$$S_S(\Omega) = (1 - \frac{\eta_e \times \sqrt{s}(\sqrt{s}-2)}{2[1+(\Omega/\Gamma)^2][(\sqrt{s}-1)^2+(\Omega/\Gamma)^2]}) . \qquad (5)$$

Figure 7 gives the quantum noise versus normalized pump power; the solid curve was calculated from Eq. (5) at analysis frequency of 35 MHz, and the dots with the error bars are the measured quantum noise of the single beam. The good agreement between the theoretical prediction and experimental measurement shows that we indeed indicated the below squeezing of the single beam.

## 4 Conclusion

In conclusion, we have investigated the intensity noise of the single beam of a pump-enhanced cw OPO, which was used to generate quantum correlated twin beams, as a function of pump power. With a triply resonant cavity, the oscillation threshold of our OPO was about 8.5±1.3 mW and the conversion efficiency was 0.72±0.02. The twin beams with power of 240mW and quantum correlation of 80% (-7.2 dB) at analysis frequency 3 MHz was generated. The relaxation oscillation frequencies, which depend on the pump power, were observed when the pump power of the OPO was from 12.5 mW to 28 mW. The intensity noise of the single beam was also investigated at a high pump power (more than four time the threshold). Although we did not obtain sub-Poissonian light, to our knowledge, we have directly indicated squeezing on a single beam of twin beams in experiment.

**Figure captions**

Fig. 1 Experimental setup for triply resonant cw OPO. D1, D2: photodiodes, PBS0, PBS1, PBS2: polarizing beamsplitter, P1: half wave plate for 532nm, P2 and P3: half wave plate for 1064nm.

Fig. 2 Output power of OPO measured as a function of pump power. The fit yields an OPO threshold of 8.5±1.3 mW and a slope conversion efficiency of $\eta = 0.72 \pm 0.02$

Fig. 3 Intensity difference noise reduction versus frequency at pump power of about 40 mW. Resolution bandwidth: 100 kHz, video bandwidth: 100 Hz.

Fig. 4 Normalized single beam noise versus frequency when pump power increases from 12.5 to 28 mW. Resolution bandwidth: 100 kHz, video bandwidth: 100 Hz.

Fig. 5 Normalized single beam noise versus frequency when pump power increases from 28 to 55 mW. Resolution bandwidth: 100 kHz, video bandwidth: 100 Hz.

Fig. 6 Intensity noise in single beam versus frequency at different pump powers. Resolution bandwidth: 100 kHz, video bandwidth: 100 Hz.

Fig. 7 Quantum noise in single beam versus normalized pump power. The solid curve represents the theoretical prediction, and the dots show the experimental results.

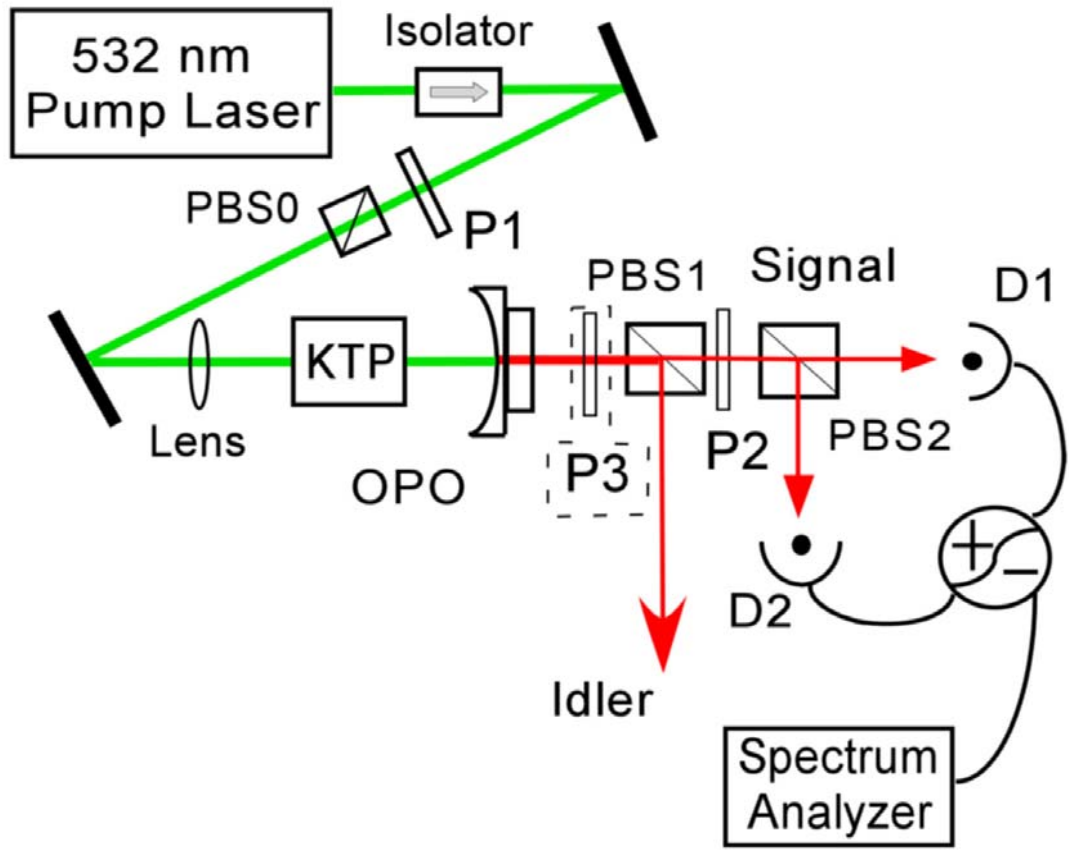

Yun Zhang et al. Fig. 1

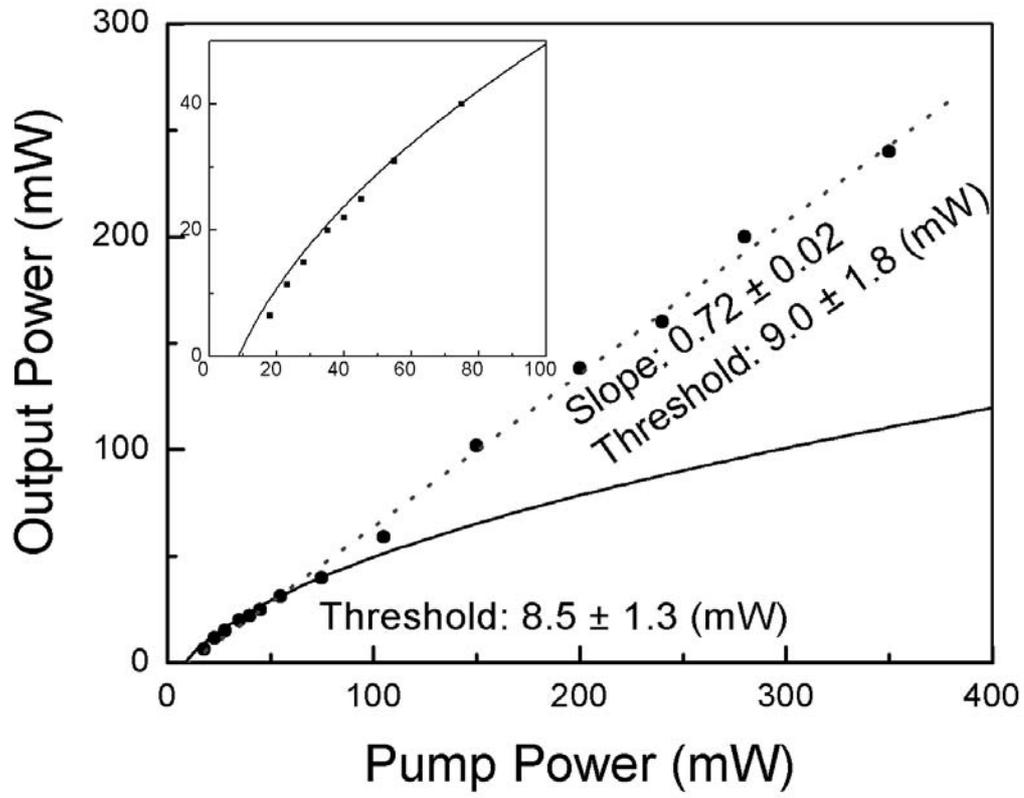

Yun Zhang et al. Fig. 2

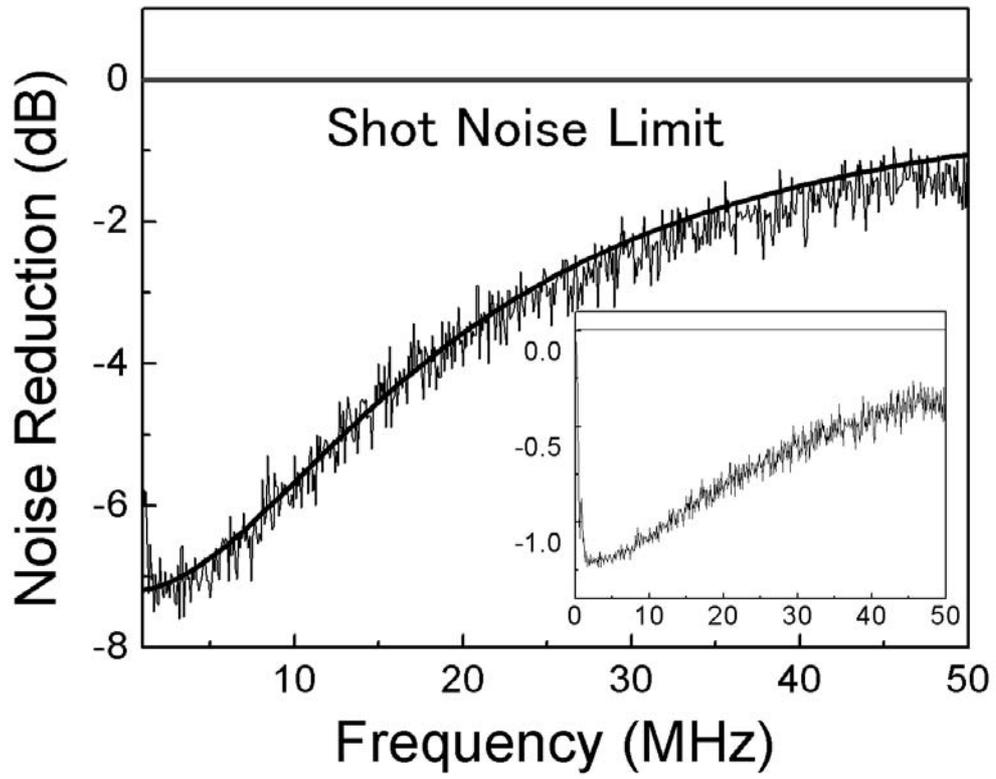

Yun Zhang et al. Fig. 3

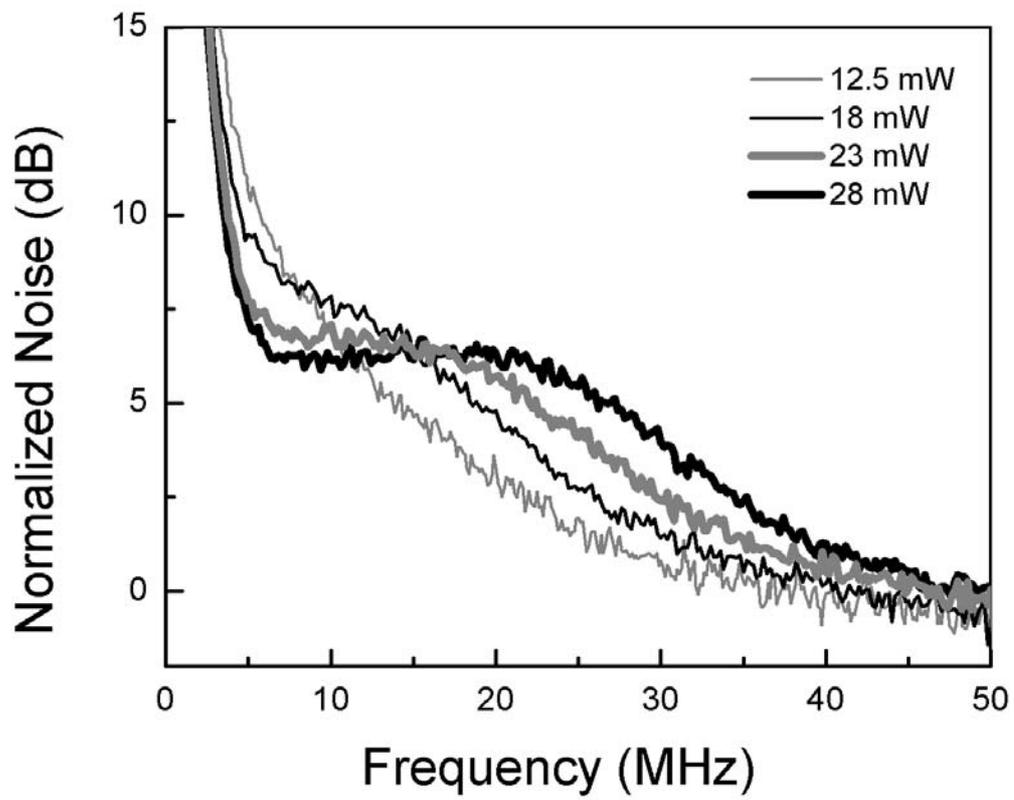

Yun Zhang et al. Fig. 4

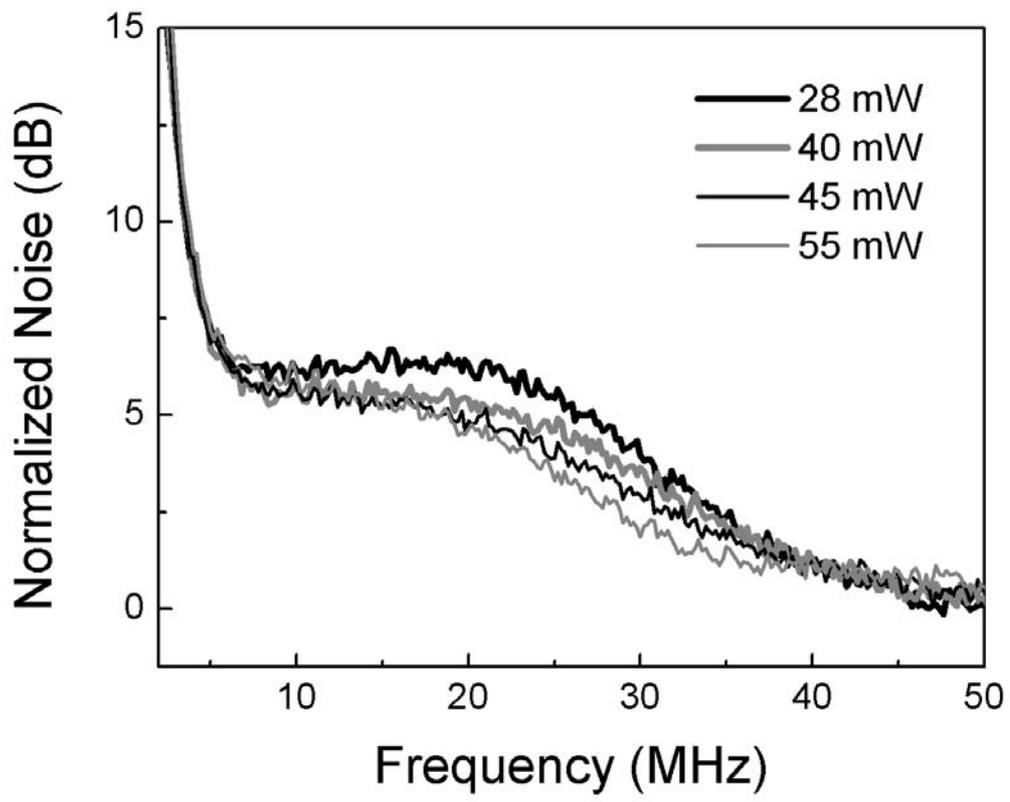

Yun Zhang et al. Fig. 5

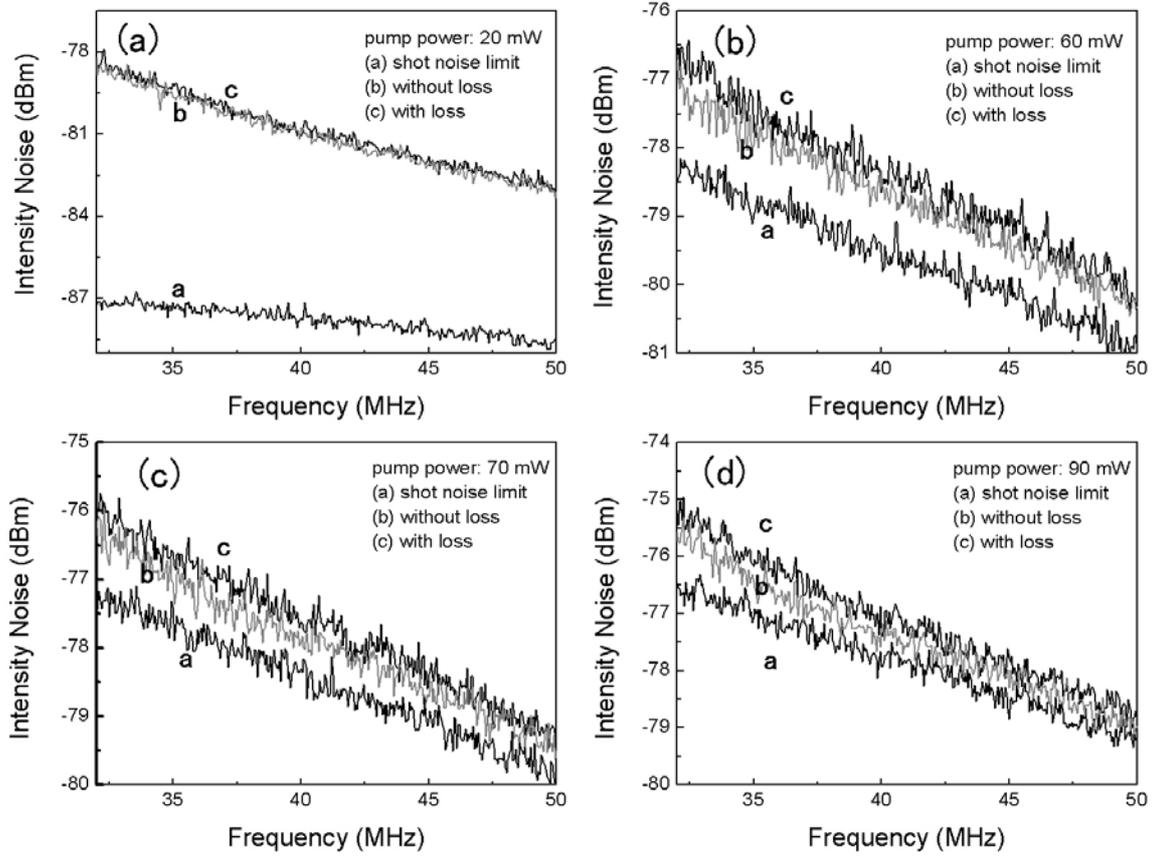

Yun Zhang et al. Fig. 6

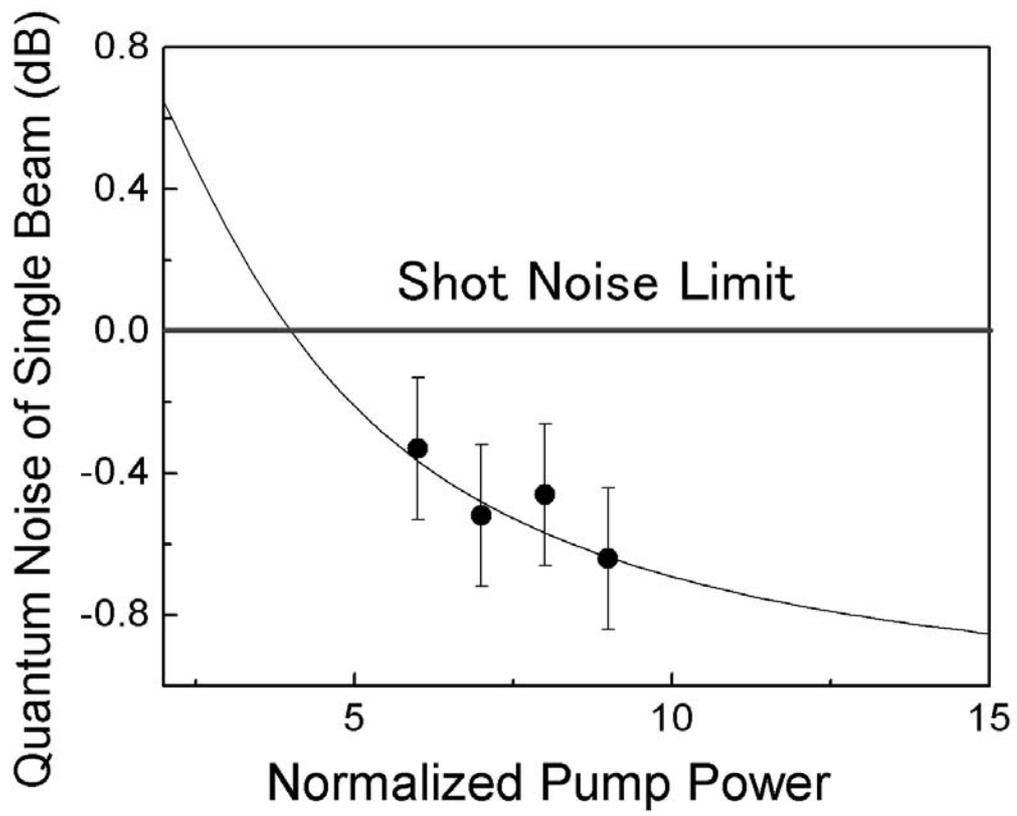

Yun Zhang et al. Fig. 7